\documentclass[pra,twocolumn, showpacs]{revtex4}

\usepackage{graphicx}
\usepackage{amsmath}

\newcommand{\ket}[1]{|#1\rangle}
\newcommand{\matelem}[3]{\langle #1|#2|#3\rangle}
\newcommand{\pf}{\rm{^{40}K}} 
\newcommand{\re}{\rm{^{87}Rb}}

\begin{document}

\title{Species-specific optical lattices}

\author{L.\ J.\ LeBlanc}  
\author{J.\ H.\ Thywissen}
\affiliation{Department of Physics, University of Toronto,  Toronto, Ontario M5S 1A7,
Canada}

\date{{\today} }


\begin{abstract}
We examine single-frequency optical schemes for species-selective trapping of ultracold
alkali-metal atoms.  
Independently addressing the elements of a binary mixture enables the creation of an optical lattice for one atomic species with little or no effect on the other.
We analyze a ``tune-in'' scheme, using near-resonant detuning to create a stronger potential for one specific element.  A ``tune-out'' scheme is also developed, in which the trapping wavelength is
chosen to lie between two strong transitions of an alkali-metal atom such that the induced dipole moment is zero for that species but is nonzero for any other.
We compare these schemes by examining the trap depths and heating rates associated with both.  
We find that the tune-in scheme is preferable for Li-Na, Li-K, and
K-Na mixtures, while the tune-out scheme is preferable for Li-Cs, K-Rb, Rb-Cs,
K-Cs and $^{39}$K-$^{40}$K mixtures.
Several applications of species-selective optical lattices are explored, including the creation of a lattice for a single species in the presence of a phononlike background, the tuning of relative effective mass, and the isothermal increase of phase space density.
\end{abstract}

\pacs{33.80.Ps, 03.75.Nt, 03.75.Ss, 39.25.+k}

\maketitle


\section{Introduction}
As the field of ultracold atoms research enters its adolescence, experiments are increasingly including more than one element or isotope. Dual-species experiments offer
possibilities for creating heteronuclear polar molecules \cite{PolarMolRev}, 
sympathetic cooling %
\cite{Cornell1997_symp,Salomon2001_LiLi, Hulet2001_LiLi,Inguscio2001_KRb},
and investigating Bose-Fermi mixtures \cite{Molmer1998,Salomon2001_LiLi, Hulet2001_LiLi,Inguscio2001_KRb} which may provide opportunities for studying boson-mediated superfluid states \cite{Heiselberg2000,Bijlsma2000}. More recently, experiments involving up to three atomic species have been implemented for sympathetic cooling of two fermionic species \cite{Dieckmann2006,three-species}.

Dually degenerate experiments have so far used external trapping potentials common to both atomic species.  A species-specific trapping potential would add a
degree of freedom to improve sympathetic cooling \cite{onofrio03}, tune effective mass, or create a lattice for one species in the presence of a background reservoir. Though careful selection of internal atomic states can provide differential magnetic trapping, optical far-off resonant traps (FORTs) and magnetostatic traps are not species-specific. 

Species-selective adiabatic potentials have been proposed \cite{Courteille2006} and demonstrated \cite{ICAP} in the case of $\re$-$\pf$, where the Land\'{e} factors  $|g_F|$ are distinct. A radio-frequency transverse field can be resonant with only one of the two species, selectively deforming its dressed potential.  Onofrio and co-workers \cite{onofrio02,onofrio03} propose using two overlapping FORTs at frequency detunings far above and below the dominant ground state transitions of both species in a two-species mixture.  The confinement of each species can be chosen independently by individually adjusting the intensity of the two beams used to create the trap.  Unfortunately, neither of these schemes lends itself to a uniform lattice potential for atoms: the radio-frequency scheme fails because it is limited to one-dimensional periodic potentials and the two-frequency balancing because lattice periodicity depends on wavelength \cite{footnote_2species}.

In this paper we discuss the generation and application of species-specific optical lattice potentials. Our motivation is the strong analogy between atoms in optical lattices and electrons in crystalline solids. Cold bosons in lattice potentials can be used to explore strongly interacting many-body physics, such as the superfluid-insulator transition \cite{Jaksch1998,Hansch02}. At sufficiently low temperatures, cold fermions in lattices \cite{Esslinger2005,Ketterle_LatticeSF2006} might be able to  address open questions about the ground state of the Hubbard Model \cite{LukinHTSC}. Unlike crystal lattices, optical lattices do not support the lattice vibrations responsible for many physical phenomena.  However, phonon mediation between neutral atomic fermions could arise in the presence of a condensed bosonic species capable of sustaining phonon-like excitations \cite{Heiselberg2000,Bijlsma2000,Viverit2002,Santamore2004}.  Although Bose-Fermi mixtures were recently loaded into optical lattices \cite{Esslinger06,Sengstock06}, the lattices confined bosons as well as fermions.  
A lattice-induced increase in the effective mass of the bosons reduces the speed of sound in the condensate, and thus the mediating effects of the phonons.  If the fermions were confined in a lattice without effect on the bosonic cloud, this mediation would be less inhibited \cite{Demler2005}. 

We discuss two approaches to species-specific optical potentials, both of which involve only a single frequency of laser light. The first approach is to tune the laser wavelength close to the atomic resonance of one species, making its induced dipole moment much stronger than that of any other atomic species present. We refer to this strategy as the ``tune-in'' (TI) scheme.
A  second approach exists for atoms, such as alkali metals, with an excited state fine structure splitting. Between the resonances of the doublet, a wavelength can be chosen such that the induced dipole moment is strictly zero. We refer to this strategy as the ``tune-out'' (TO) scheme. Both approaches allow for the creation of a species-specific optical lattice with a tunable relative potential strength between species.

In the following sections we consider the relative merits of the TI and TO schemes. 
We focus our attention on Bose-Fermi
combinations throughout, paying special attention to mixtures including $\re$ \cite{roati,Jin_BECDFG_PRA2004,Esslinger2005,SengstockBECDFG,DFG, Arimondo2005, Zimmerman2005}.
Section \ref{sec:methods} presents the methods used to calculate potential energies and heating rates of atoms in a light field. We present the TI and TO approaches in Sec.\ \ref{sec:schemes}, and quantify their relative merits. For several of the stable alkali-metal isotopes, we calculate the tune-out wavelengths, scattering and associated heating rates, and the trapping potentials 
imposed upon a second species of alkali-metal atom. 
Section \ref{sec:interactions} discusses the role of interactions, both in limiting the distinct addressability of one species and in allowing thermalization between two species.
In Sec.\ \ref{sec:applications} we discuss applications of the species-specific dipole potentials for cooling, engineering mobility, and adding phonons to optical lattices, before concluding in Sec.\ \ref{sec:conclude}.

%
\section{Methods}
\label{sec:methods}

We consider the light-atom interaction in the limit of a small excited-state fraction.
An electromagnetic field induces an electric dipole potential on a neutral atom, calculated in Sec.\ \ref{sec:dipole_calcs} which can be used as a trapping  potential for ultracold atomic samples.  
We  consider the residual effect of spontaneous emission in \S\ref{calc:heating}.

\subsection{Dipole potential}
\label{sec:dipole_calcs}
An atom in a ground state $|g\rangle$ will experience a potential shift due to coupling by the light field to the excited states $|e\rangle$. We calculate the sum of these shifts on each state $|g\rangle$, including the counter-rotating term, using second-order perturbation theory:
\begin{equation} \label{eq:lightshift}
           U_{g} = \frac{1}{2 \epsilon_{\rm 0} c} \sum_e \left[
           \frac{|\matelem{e}{\mathbf{d}\cdot\mathbf{\hat{\epsilon}}}{g}|^2}
           {\hbar ( \omega_{L}-\omega_{eg})} 
           - \frac{|\matelem{e}{\mathbf{d}\cdot\mathbf{\hat{\epsilon}}}{g}|^2}
           {\hbar ( \omega_{L} + \omega_{eg})}
           \right] I,
           \end{equation}
where $\omega_{L}$ is the laser frequency,
$\hbar \omega_{eg}$ is the energy difference between $|e\rangle$ and $|g\rangle$, 
$\mathbf{d}$ is the dipole operator, $\mathbf{\hat{\epsilon}}$ is the polarization of the light, and $I$ is the light intensity. In the case of atoms in a weak magnetic field, we use the matrix elements defined in the $\ket{F, m_{\rm F}}$ basis, where $F$ is the total angular momentum and $m_{\rm F}$ is the magnetic quantum number.

We will consider only alkali-metal atoms, which have two dominant $ns \rightarrow np$ transitions
due to the fine structure splitting. 
Using nomenclature established by Fraunhofer for the 3$^2S_{1/2}\rightarrow 3^2P_{1/2}$  and 3$^2S_{1/2}\rightarrow 3^2P_{3/2}$ transitions in sodium, we label the corresponding lines in each of the alkali metals $D_1$ and $D_2$, respectively. 
 Spin-orbit coupling splits each excited state by a frequency $\Delta_{\rm FS} = \omega_{D_2} - \omega_{D_1}$, while each ground and excited
state is further split by the hyperfine interaction $\Delta_{\rm HFS}$  and $\Delta^{\prime}_{\rm HFS}$, respectively.
 The atomic data used for Eq.~(\ref{eq:lightshift}) are the measured linewidths and line centers of the $D_1$ and $D_2$ lines, and the ground and excited state hyperfine splittings \cite{atomicdata}.
 
 Transitions to higher excited states $ns \rightarrow (n+1)p$ are neglected by our treatment. 
When detuned within $\Delta_{FS}$ of the $ns \rightarrow np$ transition, the relative magnitude of the $ns \rightarrow (n+1)p$ shift is less than $2\times10^{-5}$ for Cs and $7\times10^{-8}$ for Li.

As Eq.~(\ref{eq:lightshift}) requires a sum over several states and knowledge of individual matrix elements, it is useful to have an approximate but simpler expression for $U_{g}$. If the detunings $\Delta_{eg} = \omega_{L} - \omega_{eg}$ 
are small compared to $\Delta_{\rm FS}$, but large compared to the excited state hyperfine splitting $\Delta'_{\rm HFS}$, an approximate expression for the dipole shift is
\cite{Grimm99}
\begin{equation} \label{eq:lightshift_approx}
           U_{g} \approx \frac{\pi c^2 \Gamma}{2 \omega_0^3}\left(
           \frac{1-P g_F m_{\rm F}}{\Delta_{\rm 1}} +
           \frac{2+P g_F m_{\rm F}}{\Delta_{\rm 2}}
           \right) I,
\end{equation}
where $P = 0, \pm1$ for $\pi$, $\sigma^{\pm}$ polarization, respectively,   $g_F$ is the Land\'{e} factor, $\Delta_{1(2)}$ is the detuning from the D$_{1(2)}$ line, $\omega_0 = (\omega_{D_1} + 2 \omega_{D_2})/3$ is the line centre weighted by line strength, and $\Gamma = (\Gamma_{D_1} + \Gamma_{D_2})/2$ is the average of the $D_1$ and $D_2$ linewidths. 
Since $\Gamma_{D_2}/\Gamma_{D_1}  \approx 1+3 \Delta_{\rm FS}/\omega_0$, one can expect an accuracy between  $\pm7$\% for Cs and $\pm0.003$\% for Li.
At small detunings, we empirically find that Eq.~(\ref{eq:lightshift_approx}) deviates from Eq.~(\ref{eq:lightshift}) by  $\ge$1\% for detunings $\min\{|\Delta_1|,|\Delta_2|\}/2 \pi \lesssim 1.5 \sqrt{M}$ GHz, where $M$ is the mass in atomic units.

As an approximate form, Eq.~(\ref{eq:lightshift_approx}) neglects the counter-rotating terms. For $|\Delta| < \Delta_{\rm FS}$, the strength of the counter-rotating contribution is at most $\Delta_{\rm FS}/2 \omega_{L}$ relative to the contribution of one near-resonant dipole transition. Thus the neglected shift is at most -2\% for Cs, and -0.001\% for Li.

\subsection{Heating Rate}
\label{calc:heating}

Detuning and intensity of optical traps must be chosen with consideration of the incoherently scattered trapping light that heats the atoms.  For each state $|g\rangle$, we quantify the rate of scattering in the low saturation limit,
\begin{equation} \label{eq:scattering}
           \gamma_{g} = 
           \sum_e 
           \frac{\Gamma_{e} |\matelem{e}{\mathbf{d}\cdot\mathbf{\hat{\epsilon}}}{g}|^2}
	    {\Delta_{eg}^2
	    } I,
\end{equation}
where ${\Gamma_{e}}$ is the natural linewidth of the $g\rightarrow e$ transition, and $\Delta_{eg} = \omega_{L} - \omega_{eg}$.  The rate of scattering of photons can be converted to an average heating rate,
$H_g = \frac{2}{3} E_{R} {\gamma}_{g}$, 
where $E_{R }= \hbar^2 k^2/2m$ is the recoil energy, $k=\omega_{L}/c$, and $c$ is the speed of light.

\section{Species-specific dipole trapping schemes}
\label{sec:schemes}
\begin{table*}[tb!] \center
\begin{ruledtabular}
\begin{tabular}{c c c c c c c c}
 \multicolumn{2}{c}{} &\multicolumn{5}{c}{Spectator} \\

Target & $\rho$ & $^7$Li  & $^{23}$Na  &$^{39}$K & $^{87}$Rb & $^{133}$Cs \\
\hline  
$^6$Li &$\infty$ & 0.00134  & 7.77$\times10^{-4}$ & -0.0381& 
-1.20& -8.45 \\

& 100& 2.66$\times10^{-5}$& 0.281 & 0.220 & 0.239 & 0.347\\

& 10&   3.92$\times10^{-6}$& 2.54 &  2.30 & 2.49 & 3.57\\

 $^{40}$K & $\infty$ & 4.28$\times10^{-7}$ & 5.77$\times10^{-4}$ &  0.188 &
-9.03 & -25.8 \\

& 100 & 3.05  & 5.78 &  3.64$\times10^{-5}$& 0.251 & 1.39\\

& 10 &29.2 &53.0& 4.65$\times10^{-4} $& 3.27 & 18.8\\

\end{tabular}
\end{ruledtabular}
\caption{Sustainability $\textsf{s}$ of two-species mixtures for tune-out and tune-in schemes, in units of
seconds, using Eq.~(\ref{eq:sustainability}).  The second column indicates the selectivity.  Rows with $\rho = \infty$  are calculated using the TO scheme while rows with $\rho = 100$ and $\rho = 10$ use the TI scheme. For the lighter spectators, the TI scheme has higher $|\textsf{s}|$.  The heaviest elements and isotope mixtures favor the TO scheme.}
\label{tab:mixtures}
\end{table*}

It is not surprising that optical traps can be species-specific, given that optical resonances are unique to atomic elements and isotopes. However, most species-specific optical traps, such as magneto-optical traps, are tuned to within a few linewidths of resonance, which is incompatible with quantum degenerate ensembles. At low temperatures and high density, any gain in trap depth close to resonance must be balanced against the heating due to unwanted light scattering (Sec.\ \ref{calc:heating}).

In the subsections below we consider  two-species mixtures. The goal is to apply a dipole force to the ``target'' species while inducing as little potential as possible on the second species, which we will call the ``spectator''.  We define the ``selectivity'' as 
\begin{equation} \label{eq:selectivity}
\rho = \left| \frac{U_{t}}{U_{s}} \right|,
\end{equation}
where $U_{t,(s)}$ is the potential induced on the target (spectator). 

As discussed in Sec.\ \ref{sec:methods}, in the low saturation limit, both the induced dipole potential and the heating rate are proportional to intensity. We define the intensity-independent ratio
\begin{equation}  \label{eq:sustainability}
\textsf{s} = \frac{U_{t}}{H_{t} + H_{s}}
\end{equation}
to be the ``sustainability'', where $H_{t(s)}$ is the heating rate of the target (spectator).  The absolute value of \textsf{s} sets the scale for possible trapping time \cite{trap_time_footnote}. The laser frequency will be chosen to maximize both $\rho$ and $|\textsf{s}|$. 

\subsection{The tune-in scheme}
\label{sec:tunein}

The simplest selective potential is one in which the laser is tuned close to a resonance of the target (see Fig.~\ref{fig:crossover}). Since scattering rate is inversely proportional to the square of the detuning, we consider only heating of the target such that \textsf{s}$_{\rm TI} \rightarrow U_{t}/H_{t}$. Considering Eq.~(\ref{eq:lightshift_approx}) in the limit $|\Delta_t| \ll  |\Delta_s|$, a simple estimate is
\begin{equation} \label{sustainability_TI}
\textsf{s}_{\rm TI} \approx \left( \frac{3 \hbar}{2 E_{R,t} \Gamma_{t}} \right) \Delta_{t}.
\end{equation}
The choice of detuning $\Delta_{t}$ will depend upon the desired selectivity. Assuming that the spectator is far-detuned, $\Delta_{s} \approx (\omega_{0,t} - \omega_{0,s})$ and  
\begin{equation} \label{selectivity_TI}
\rho_{\rm TI} \approx \left( \frac{2 \Gamma_{t}}{3 \Gamma_{s}} 
\frac{\omega_{0,s}^3 |\omega_{0,t} - \omega_{0,s}|}{\omega_{0,t}^3}
\right) \frac{1}{|\Delta_{t}|}.
\end{equation}
Together, Eqs.~(\ref{sustainability_TI}) and (\ref{selectivity_TI}) explicitly show the opposing dependence on $\Delta_{t}$ and the necessary trade-off between selectivity and sustainability.  

Table \ref{tab:mixtures} shows the sustainability for bosonic species as spectators and
fermionic species as targets, calculated using Eqs.\ (\ref{eq:lightshift}) and (\ref{eq:scattering}) for tune-in selectivities
of 100 and 10. The inverse scaling predicted by Eqs.\ (\ref{sustainability_TI}) and (\ref{selectivity_TI}) is
observed: $\textsf{s}_{\rm TI}$ drops by approximately a factor of ten when $\rho_{\rm TI}$ is
increased by a factor of ten. The product of $\textsf{s}_{\rm TI}$ and $\rho_{\rm TI}$ in Table \ref{tab:mixtures}
varies between 22 s and 36 s for $^6$Li mixtures and between 140 s and 530 s for
$^{40}$K mixtures, whereas Eqs.\ (\ref{sustainability_TI}) and (\ref{selectivity_TI}) would predict a ranges of  8 s to 21 s and  64 s to 132 s, respectively, excluding isotope mixtures from the comparison. Finally, we note  that
$^{23}$Na-$^{40}$K is the optimal tune-in mixture.

\subsection{The tune-out scheme}
\label{sec:tuneout}

\begin{figure}[tb!]
\includegraphics[width=3.25in]{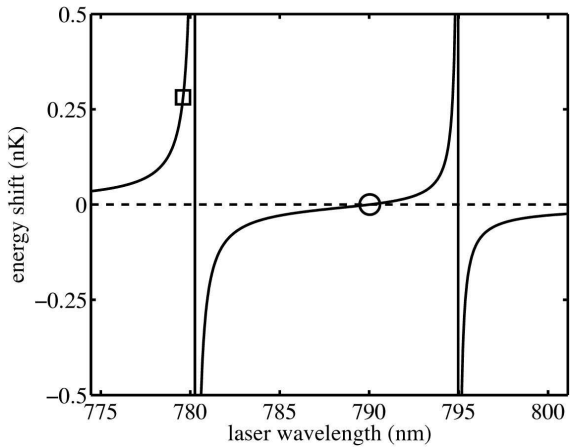}
\caption{Energy shift as a function of wavelength for $\re$ in the $\ket{F,m_F}$ =
$\ket{2,2}$ state, under linear polarization, for 1 mW/cm$^2$.  This general structure will arise for each of the alkali-metal elements, with the divergences located at the $D_{1}$ and $D_{2}$ lines.  The tune-in scheme for a $\re$  target is indicated by the square marker on the blue-detuned branch of the potential energy curve.  The round marker indicates the position of the tune-out wavelength, where the energy shift is zero. Here, $\re$ is the spectator.}
\label{fig:crossover}
\end{figure}
%

\begin{table}[t!] \center
\begin{ruledtabular}
\begin{tabular}{c l c c c} 
Polarization & $|F, m_F\rangle$ & $\lambda_{\rm TO}$ &
${\gamma}_{\rm sc}/I$
\\
 & & (nm) & (cm$^2$/mJ) \\ 
\hline 

        $\pi$   & $|2,\pm2\rangle$ & 790.05 & 9.0$\times$10$^{-6}$ \\ 
     & $|2,\pm1\rangle$ & 790.06 &  \\ 
    & $|2,0\rangle$ & 790.06 &  \\ 
     & $|1,\pm1\rangle$ & 790.06 &  \\ 
     & $|1,0\rangle$ & 790.05 &  \\
     
\\
   Linear     & $|2,\pm2\rangle$ & 790.04 &  9.1$\times$10$^{-6}$\\ 
     & $|2,\pm1\rangle$ & 790.04 &  \\ 
    & $|2,0\rangle$ & 790.03 &  \\ 
     & $|1,\pm1\rangle$ & 790.03 &  \\ 
     & $|1,0\rangle$ & 790.04 &  \\     

\\
 $\sigma^-$    & $|2,2\rangle$ & 785.14 & 9.1$\times$10$^{-6}$ \\ 
     & $|2,1\rangle$ & 787.59 & 8.1$\times$10$^{-6}$ \\ 
     & $|2,0\rangle$ & 790.06 & 9.0$\times$10$^{-6}$ \\  
     & $|2,-1\rangle$ & 792.52 & 14.4$\times$10$^{-6}$ \\ 
     & $|2,-2\rangle$ & (none) & - \\ 
     & $|1,1\rangle$ & 792.53 & 14.4$\times$10$^{-6}$ \\ 
     & $|1,0\rangle$ & 790.06 & 9.0$\times$10$^{-6}$ \\ 
     & $|1,-1\rangle$ & 787.59 & 8.1$\times$10$^{-6}$ \\

\\ 
     $\sigma^+$    & $|2,2\rangle$ & (none) & - \\ 
     & $|2,1\rangle$ & 792.52 & 14.4$\times$10$^{-6}$ \\ 
     & $|2,0\rangle$ & 790.06 & 9.0$\times$10$^{-6}$ \\ 
     & $|2,-1\rangle$ & 787.59 & 8.1$\times$10$^{-6}$ \\ 
     & $|2,-2\rangle$ & 785.14 & 9.1$\times$10$^{-6}$ \\ 
     & $|1,1\rangle$ & 787.59 & 8.1$\times$10$^{-6}$ \\ 
     & $|1,0\rangle$ & 790.06 & 9.0$\times$10$^{-6}$ \\ 
     & $|1,-1\rangle$ & 792.53 & 14.4$\times$10$^{-6}$ \\

\end{tabular}
\end{ruledtabular}
\caption{Tune-out wavelengths and scattering rates in $^{87}$Rb, for select polarizations and all ground states. Two states have no tune-out wavelength, because the $D_{\rm 1}$ line has no F=3 excited state.}
\label{tab:Rb87}
\end{table}

\begin{table}[tb!] \center
\begin{ruledtabular}
\begin{tabular}{ c c c c c} 

Element & $|F, m_F\rangle$ & \multicolumn{2}{c}{$\lambda_{\rm TO} $  (nm) }&
$\mathit{\gamma}_{\rm sc}/I$
\\
 & &Eq.~\ref{eq:lightshift} & Eq.~\ref{eq:lightshift_approx} &  (cm$^2$/mJ) \\
\hline 
 
 $^6$Li  & $|\frac{3}{2}, \frac{3}{2}  \rangle$ &  670.99 & 670.99 & 2.8\\

 $^7$Li  & $| 2, 2  \rangle$ &  670.97 & 670.97 & 2.4\\

$^{23}$Na  & $| 2, 2 \rangle$ & 589.56 &  589.56 & 2.0$\times10^{-3}$\\

$^{39}$K & $|2,2  \rangle$ & 768.95  & 768.95 & 1.4$\times10^{-4}$ \\

 $^{40}$K & $|\frac{9}{2},\frac{9}{2}  \rangle$ & 768.80 & 768.80 & 1.7$\times10^{-4}$ \\

 $^{87}$Rb & $| 2, 2 \rangle$ & 790.04 & 790.01 & 9.1$\times10^{-6}$\\
 
 $^{133}$Cs & $| 4, 4 \rangle$ & 880.29 & 880.06 & 1.5$\times10^{-6}$\\

\end{tabular}
\end{ruledtabular}
\caption{Tune-out wavelengths and scattering rates for various elements. We have assumed linear
polarization and stretched states and show a comparison of Eqs.\  (\ref{eq:lightshift}) and (\ref{eq:lightshift_approx}).}
\label{tab:lambdac}
\end{table}

The tune-out wavelength scheme exploits the characteristic doublet structure of the alkali-metal atoms.   By choosing
a wavelength that lies between the two strongest transitions,
the large negative energy shift of the $D_2$ line is balanced against the large positive energy shift of the $D_1$ line (see Fig.~\ref{fig:crossover}). This atom becomes a spectator, while any other species feels the shift induced by the laser and becomes a target. Since the potential shift on the spectator can be zero, the selectivity of the tune-out approach is infinite.

The laser frequency, $\omega_{\rm TO}$,  at which $U_{s} = 0$, is determined numerically using Eq.~(\ref{eq:lightshift}) with all $\ket{F, m_ {\rm F}}$ excited states in the $D_{\rm 1}$ and $D_{\rm 2}$ manifolds. Table~\ref{tab:Rb87} shows the 
tune-out wavelength for all $\re$ ground states with  $\pi$, linear \cite{poln_footnote}, $\sigma^+$, and $\sigma^-$ polarizations. An approximate expression for this wavelength can be derived from Eq.~(\ref{eq:lightshift_approx}), giving
\begin{equation} \label{eq:lambda_TO}
\omega_{\rm TO} = \omega_0 - \frac{1 + P g_{\rm F} m_{\rm F}}{3} \Delta_{\rm FS}.
\end{equation}
For $\re$, Eq.~(\ref{eq:lambda_TO}) predicts tune-out wavelengths of 785.10 nm, 787.54 nm, 790.01 nm, and 792.49 nm for $P g_{\rm F} m_{\rm F} =$ -1, -0.5, 0, and 0.5, respectively. Comparing these values to the results of Table~\ref{tab:Rb87}, we see the that approximations are accurate to 0.04 nm or better. Equation~(\ref{eq:lambda_TO}) also gives $\omega_{\rm TO}=\omega_{\rm D_1}$ for the case of $P g_{F} m_{F} = 1$, which is inconsistent with the assumption that $|\Delta_1| \gg \Delta_{\rm HFS}$. In fact, since $Pg_{F}m_{F} = 1$ corresponds to a dark state with respect to the D$_1$ excitation, there is no tune-out wavelength for this case.

We note that in Table~\ref{tab:Rb87} the tune-out wavelength for linear
polarization is nearly independent of the choice of ground state, in contrast to  $\sigma^+$ or
$\sigma^-$ polarizations. Given this independence, we calculate the tune-out wavelengths using Eq.~(\ref{eq:lightshift}) for several of the common alkali-metal isotopes in their stretched ground states under linear polarization (Table \ref{tab:lambdac}) \cite{footnote_polnaxes}.

Tables \ref{tab:Rb87} and \ref{tab:lambdac} also give the scattering rates per unit intensity (scattering cross-section) for the spectator species at the tune-out wavelength. 
Due to the large dispersion of fine splitting among the alkali metals, scattering cross-sections vary by over six orders of magnitude. To understand the implications for trapping, we need to consider the sustainability, $\textsf{s}$, of various pairs of atomic species. 
Table \ref{tab:mixtures} shows $\textsf{s}_{\rm TO}$ in the same mixtures for which $\textsf{s}_{\rm TI}$ is calculated.  Spectators with larger scattering cross-sections at $\lambda_{TO}$ have lower sustainability.
In addition to these data, we note that the $^{133}$Cs-$^{87}$Rb
spectator-target combination gives the highest possible sustainability among alkali metals in the tune-out scheme: $\textsf{s}=34$\,s at a tune-out wavelength of $\lambda_{\rm TO} = 880.29$ nm (not shown in tables).

As an example, consider making a lattice
for  $\pf$ only, leaving $\re$ unaffected by the lattice and confined only by a background magnetic trap or FORT.  If we consider the mixture in a three-dimensional
lattice of arbitrary depth,  the $\pf$ potential shift is $1.54 \times10^{-5}\mu K\times[I(\mathrm{mW/cm^2})]$; with beams of 100$\mu$m waist, the potential shift is 98 nK/mW. If we require the target to experience a lattice that is  $8E_{R}$ deep \cite{8ERnote}, where $E_{R}$ is the recoil energy, we find a $\re$ heating rate of \mbox{98~nK/s}.

%
\subsection{Discussion: tune-in vs.\ tune-out}

With the data of Table~\ref{tab:mixtures}, we can evaluate the practicality of both the tune-in and tune-out schemes for Bose-Fermi mixtures of neutral alkali metal atoms.  Since the scattering rate in the tune-out scheme can be smaller for elements with larger fine structure splittings, this approach is better suited to more massive elements \cite{FS_footnote}. In particular, the tune-in scheme is preferable for Li-Na, Li-K, and K-Na mixtures, and for applications requiring selectivity of less than 10:1. The tune-out scheme is preferable for Li-Cs, K-Rb, and K-Cs mixtures when the selectivity required is greater than 10:1, and for Li-Rb mixtures at selectivity of greater than 20:1.  

An isotope mixture of potassium could be compatible with the tune-out scheme, where \textsf{s} $>$ 100 ms. Isotope-specific manipulation within a lithium mixture is less practical due to sustainabilities of 1 ms or less.  

Other factors may also influence whether the tune-in or tune-out approach is preferred.  For experiments with time scales that are slow compared to the thermalization rate (discussed further in Sec.\ \ref{sec:rethermalization}), it may be preferable to heat the minority species and allow for sympathetic cooling.
If the reservoir of spectator atoms is large, the tune-in scheme might be preferred since extra energy due to near-resonant heating would be transferred to the reservoir.
For experiments on time scales fast compared to the thermalization rate, the tune-out scheme might be preferred even if \textsf{s} is smaller, since spectator heating will not affect the target.

Finally, we note that if a third species is included, it will play a spectator role in the tune-in scheme and a  target role in the tune-out scheme.

\subsection{Lattice confinement effects}
\label{sec:confinement}

In Secs. A-C we have discussed the ratio of induced potential to heating rate as a local quantity. However, the spatial dependence of intensity in an optical lattice means that moving atoms may not sample all laser intensities uniformly and the heating rate should be modified accordingly. A more relevant characterization of trapping and dissipation is
\begin{equation} \label{s_global}
\textsf{s}_{\rm global} 
= \frac{U_{\rm t}^{\rm pk} }{\langle H_{t} \rangle + \langle H_{s} \rangle},
\end{equation}
where the brackets indicate expectation values and $U_{\rm t}^{\rm max}$ is the maximum potential depth of the target.  In the case where only one heating rate dominates, we define a quantity $\beta$ to represent the enhancement of sustainability due to the lattice: 
\begin{equation}
\label{eq:beta}
\beta_{\rm TO, TI} \equiv  \frac{ \textsf{s}_{\rm global} }{\textsf{s}}
\rightarrow \frac{ I_0 }{\langle I \rangle}_{s,t},
\end{equation}
where the expectation value is taken for the spectator species in the tune-out case, and for the target in the tune-in case.  In a one-dimensional lattice, the intensity  is  $I(x) = I_0 \mathrm{sin}^2(k x)$, where $k = 2 \pi / \lambda_{L}$.

For the tune-out scheme, neither the position nor the motion of the spectator is coupled to the standing wave intensity.  We can assume that all intensities are equally sampled and use the average intensity; ${\langle I \rangle}_{s,t} =I_0/2$ to give $\beta_{\rm TO} = 2$.

\begin{figure}[tb!]
\includegraphics[width=8.25cm]{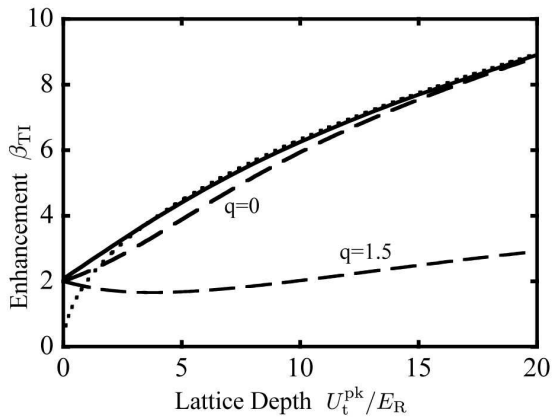}
\caption{A standing wave of laser light increases
 the ratio of confinement depth to scattering rate, modifying the sustainability \textsf{s} calculated for uniform intensity. The enhancement $\beta_{\rm TI}$ is shown as a function of $\eta$ (trap depth) for a localized state (solid line), a $q = 0$ Bloch state in the lowest band, and a $q = 1.5 \hbar k$ Bloch state in the first excited band (both dashed). At high lattice depths, both wavefunctions in the lowest band approach the tight binding limit (dotted line). For comparison, $\beta_{\rm TO}=2$ for all lattice depths.}
\label{fig:blue_adv}
\end{figure}

For the tune-in scheme, the target is the species for which the heating rate is dominant and for which the lattice enhancement of the sustainability must be taken into account. We first consider $\beta$ in the case of a localized atom. Classically, we expect deeply bound states of a \emph{blue}-detuned lattice to avoid the intensity maxima where incoherent scattering is high, such that the intensity seen by the atom is less than the average. Heating is therefore reduced and $\beta_{\rm TI}$ is increased.  For an atom localized to a single site by interactions, we can use Wannier states to calculate the expectation value of the intensity (see Fig.~\ref{fig:blue_adv}). For lattice depths $\eta = U_{\rm t}^{\rm pk}/E_{\rm R} \gg 1$, these states are approximately harmonic oscillator states with oscillator frequencies $\omega_{\rm ho} = 2 \sqrt{E_{R} U_{t}}/\hbar$. In this case,
$\beta_{\rm TI} \approx 2 \sqrt{\eta}$; in an arbitrary number of  dimensions, $d$,  the advantage is increased as $\beta_{\rm TI} \approx 2^d \eta^{d/2}$ for $\eta \gg 1$.

Considering that classical oscillators spend more time at turning points than at  potential minima,  blue tune-in lattices may not always have $\beta_{\rm TO}>\beta_{\rm TI}$. For a delocalized  $q=1.5 \hbar k$ Bloch function (in the first excited band), where $q$ is the quasi-momentum, we show $\beta$ versus $\eta$ in Fig.~\ref{fig:blue_adv}. For $\eta < 10$, confinement effects give $\beta_{\rm TI}<2$. 

\section{Interactions between species}
\label{sec:interactions}

\subsection{Mean field interactions} 
Interactions between elements may couple the environment felt
by the target to the spectator, spoiling the species-specific potential shaping of both schemes.
 If target atoms are trapped in a lattice in the presence of a background spectator species, a
periodic interaction potential could arise for the spectator due to its interaction with the periodically modulated density of the target. 
A mean-field approach is used to make a simple estimate of the magnitude of this effect.  The interaction potential of the spectator due to the target is given by $U_{\mathrm{int},s} = (4 \pi \hbar^2
a_{st} /\mu_{st}) n_t$, where $a_{st}$ is the scattering length between species, $n_{t}$ is the density of the target, 
and $\mu_{st} = 2 m_{s} m_{t}/(m_{s}+m_{t})$ is the reduced mass.  As ${U_{t}}/{U_{\mathrm{int},s}} \rightarrow 1$, 
interaction effects begin to degrade selectivity.  Estimating the three-dimensional density of the target at the centre
of a single lattice site using the harmonic oscillator ground state with frequency $\omega_{\rm
ho}$ and assuming one atom per site, we find an interaction-limited selectivity 
\begin{equation}
\label{eq:meanfield}
\rho_{\rm max}  \equiv \bigg|\frac{U_{\rm t}}{U_{\mathrm{int},s}}\bigg| = \frac{\hbar
\mu_{st} \eta^{1/4}}{32 \pi |a_{st}| m_{t}^{3/2} E_{R}^{1/2}}.
\end{equation}
In the case of the  $^6\mathrm{Li}$-$\re$ target-spectator mixture, the interaction-limited selectivity is $\rho_{\rm max} = 3.1\eta^{1/4}$ \cite{Zimmerman2005}, while for the $\pf$-$\re$ target-spectator mixture, it is $\rho_{\rm max} = 0.22\eta^{1/4}$.
To be in a regime where interaction effects might be ignored, we require $\rho_{\rm max} \gg 1$ which gives $\eta \gg 0.01$ for $^6\mathrm{Li}$-$\re$ and $\eta \gg 400$ for $\pf$-$\re$.   While for the lithium target, this condition is quite reasonable, the large lattice depths required for potassium would completely localize the atoms to individual sites and prevent the exploration of interesting tunneling-driven physics. 
Overcoming this interaction limitation and achieving the high selectivity discussed in Sec.\ \ref{sec:schemes} may be possible by tuning a magnetic field to a value where $a_{\rm RbK}
\approx 0$ near a Feshbach resonance \cite{Jin_Feshbach}.   For instance, with depth  $\eta = 8$ and $\rho_{\rm max} \ge 10$, the scattering length between species is limited to $|a_{\rm st}| \le 7.5a_{\rm 0}$, where $a_{\rm 0}$ is the Bohr radius.

Where species selectivity is not the goal, this interaction-induced periodic potential for a second species could be used alternative lattice potentials.  Such potentials are non-sinusoidal, do not involve a Stark shift, and may have a dynamic structure if target atoms are mobile.  The strength of this induced potential could be controlled through the interaction strength, such as at a Feshbach resonance, as described above.

\subsection{Inter-species thermalization}
\label{sec:rethermalization}

An understanding of the thermalization between species is important when considering heating or cooling in the trap.  The rate at which energy is transferred between the species will be relevant in setting the time scales on which adiabatic experiments can take place.    Thermometry is also possible if there is good thermalization between species; a high density target could be confined within a species-specific lattice  while the spectator remains extremely dilute and thus at lower quantum degeneracy, where temperature is more easily measured.  Conversely, thermal isolation could be useful in shielding one species from the spontaneous heating in the other.

In the classical limit, the thermalization rate is proportional to the collision rate of the atoms in the trap, given by $\gamma_{coll} = n \sigma v$, where $n$ is the overlap density, $\sigma$ the scattering cross-section, and $v$ the relative velocity between species. Random collisions act to equilibrate the system and the rate of rethermalization is $\gamma_{therm} \approx \gamma_{coll}/2.7$ \cite{retherm_coll}.

For a degenerate mixture of bosons and fermions, the classical picture of scattering breaks down and the rate of  thermalization  decreases as the fermionic system becomes more degenerate.  An estimate of the sympathetic cooling of a uniform system in the degenerate regime using the quantum Boltzmann equations gives the rate of change in the degeneracy \cite{Carr_2004}:
\begin{equation} \label{eq:thermalization}
\frac{d}{dt}\left(\frac{T}{T_F}\right) = -\frac{6 \zeta(3)}{\pi^2} \gamma'_{coll} \left 
                        (\frac{T}{T_F}\right)^2,
\end{equation}
where $\gamma'_{coll} = (3/8)n_{B} \sigma  v_{F}$ is the collision rate between species, with the
scattering cross-section $\sigma = 4\pi a_{BF}^2$, Fermi velocity, $v_{\rm F} = \hbar
 (6 \pi n_{\rm F})^{1/3}/m_{\rm F}$, $\zeta(3) \approx 1.20206$ is the Riemann zeta function and $T_{\rm F}$ is the Fermi temperature.  Though other assumptions \cite{footnote_assume} of this treatment are not valid for the systems we consider in this paper,  we use this expression to determine an order of magnitude for the thermalization rate. 

Taking boson density $n_B = 1\times10^{14}$ cm$^{-3}$,  fermion density $n_F = 1\times10^{13}$ cm$^{-3}$, and using the mass for $\pf$,  we calculate that $d/dt(T/T_F) \approx -350 s^{-1} (T/T_F)^2$, which gives a temperature relaxation time  of $ \approx 30$ ms at $T = 0.1 T_F$.  
This is an order of magnitude larger than the the classical expectation for the rethermalization time of $\approx 2$ ms for particles moving at the Fermi velocity.   Thus for rapid experiments in the deeply degenerate regime, thermal contact is essentially broken, allowing, for instance, the target to be unaffected by the heating of the reservoir in the tune-out scheme.


\section{Applications of species specific trapping}
\label{sec:applications}
\subsection{Isothermal phase space increase}

Two-species mixtures can be used to realize various cooling schemes.  For example, dark state cooling by superfluid immersion is discussed in Ref.\ \cite{Zoller2006}.  We present two simple cooling scenarios in which the presence of an uncompressed
spectator allows the target species to be compressed with negligible temperature increase but significant improvement
in phase space density. 
In both cases, target atoms are first compressed isothermally, spectator atoms are then removed from the trap, and finally, the target atoms are decompressed adiabatically.

We consider a species-selective single well dipole trap  and a species-selective lattice.  A closed cycle is described for both scenarios with atoms beginning and ending in a FORT.
We assume that the heat capacity ratio between spectator and target is infinitely large and that thermalization in the spectator and between the spectator and target
species is faster than any other time scale considered.  The latter assumption may restrict fermion cooling to the non-degenerate regime (see Sec.\ \ref{sec:rethermalization}).  

\begin{figure}[tb!]
\includegraphics[width=3.25in]{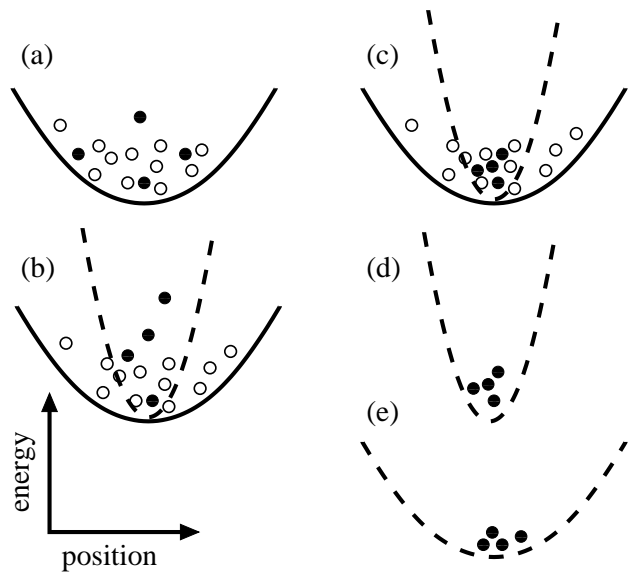}
\caption{A cooling procedure using species-specific trapping; one-dimensional trap shape represents three-dimensional trap-averaged shape. The solid line represents a FORT, the dashed line the species-specific trap; open circles are the spectator species and closed circles are the target.  (a)~Two species are trapped in a FORT;  (b)~the species-selective beam is turned on, compressing and heating the target species; (c)~the target species rethermalizes with the spectator; (d)~the spectator is removed; (e)~the target is adiabatically decompressed to a lower temperature and transferred to a FORT.}
\label{fig:cooling}
\end{figure}

In the first scenario, schematically represented in Fig.\ \ref{fig:cooling},  a single species-specific beam crosses a  FORT.  Adiabatic cooling
reduces temperature in proportion to the ratio of average trapping frequencies, that is, $T_{\rm i}/T_{\rm
f} ={\omega_{\rm i}}/{\omega_{\rm f}}$, where $i,f$ indicate ``initial'' and ``final''.   As an example,
consider $\pf$-$\re$ and $^6\mathrm{Li}$-$\re$ mixtures, confined by a 1064 nm, 500 mW single-beam FORT with a $1/e^2$
radius of 20 $\mu$m and a corresponding trap-averaged harmonic oscillator frequency ${\omega_{\rm
i}} = 2\pi\times 540$ Hz for $\pf$ and ${\omega_{\rm
i}} = 2\pi\times 950 $ Hz for $^6\mathrm{Li}$.  A 500 mW, 50 $\mu$m waist beam at $\lambda_{\rm TO}$ (see Table
\ref{tab:lambdac}) is turned on perpendicular to the FORT to compress the fermions in a trap with
frequency $\bar{\omega_{\rm f}} = 2\pi\times 3040$ Hz for $\pf$ and $\bar{\omega_{\rm f}} = 2\pi\times 3810 $ Hz for lithium.  After providing sympathetic cooling during
the compression of the target atoms,  the rubidium is ejected from the trap by temporarily removing
the FORT or by using a resonant pulse of light. The species-specific trap is then adiabatically
ramped down and turned off, leaving the fermions in the FORT at a temperature approximately 5.7 (4.0) times colder than
when they began for potassium (lithium). Though this is a modest change in temperature, the phase space density in a harmonic trap is proportional to the inverse cube of the temperature, indicating a factor of 180 increase in phase space density for potassium and a factor of 60 for lithium.  

In the second scenario, we consider a 3D lattice created by the tune-out wavelength \cite{footnote1}.  The
target-specific lattice is ramped on until peak lattice intensity is reached.  The spectator is
evaporatively cooled and ejected by reducing the spectator trap depth. The lattice is then ramped
back down isentropically, leaving the target in the initial trap with an entropy and temperature
limited by the ``plateau entropy'' discussed in \cite{Blakie}. As shown there, a target of
fermions with unity filling has an entropy plateau of zero, which would suggest no lower limit to the achievable temperature.

An important  limitation of these schemes will be the competition between adiabaticity
 and heating.  For an ideal gas, the condition of
adiabaticity requires any changes to take place in times longer than the inverse of the smallest
trap frequency.  Using the numbers given in the $\pf$-$\re$ example of the crossed dipole trap in the tune-out scheme, we find that  adiabaticity requires a relaxation time of 35 ms, at an
intensity of 3.5$\times 10^7$ mW/cm$^2$, yielding a heating of 1.3 $\mu$K during this time, which sets a
lower bound for the temperature attainable in this scheme.   Another possible limitation of the cooling schemes is the interspecies thermalization, which limits the speed of the isothermal step (see discussion in \S \ref{sec:rethermalization}).

\subsection{Phonons}

Unlike crystal lattices of solids, optical lattices do not support phonons.  Since these quasi-particles play a leading role in the physics of condensed matter, it is of interest to introduce phononlike excitations into a system of ultracold atoms in an optical lattice.  The boson-mediated interaction between fermions has been studied  in both uniform \cite{Heiselberg2000,Bijlsma2000,Viverit2002,Santamore2004}  and lattice systems \cite{Lewenstein2004,Roth2004,Demler2005}. 
Little theoretical work has been done on a system in which only the fermions are 
confined to the lattice \cite{Klein_ICAP}.  Here, the background bosons are free to interact with the fermions
and with one another.  If the bosons are degenerate, the condensate can sustain phonon-like excitations and allow for boson-mediated interactions between fermions on spatially separated lattice sites.

For the phonons in the condensate to play a role in mediating interactions, their spatial extent must exceed the lattice spacing.  The healing length of a uniform bose condensate, $\xi = (8na)^{-1/2}$ where $n$ is the condensate density and $a$ is the scattering length between species,  sets the relevant length scale.  Species-specific trapping allows the superfluid bosons to remain at low density while fermions are tightly bound in the lattice, thereby maximizing the range of the mediated interaction.  
A finite selectivity does not prevent mediation of interactions, since the bosons remain superfluid at depths less than the Mott-insulator superfluid transition \cite{Hansch02}, permitting both tune-in and tune-out schemes to be used for this application.   

\begin{figure}[tb!]
\includegraphics[width=8.25cm]{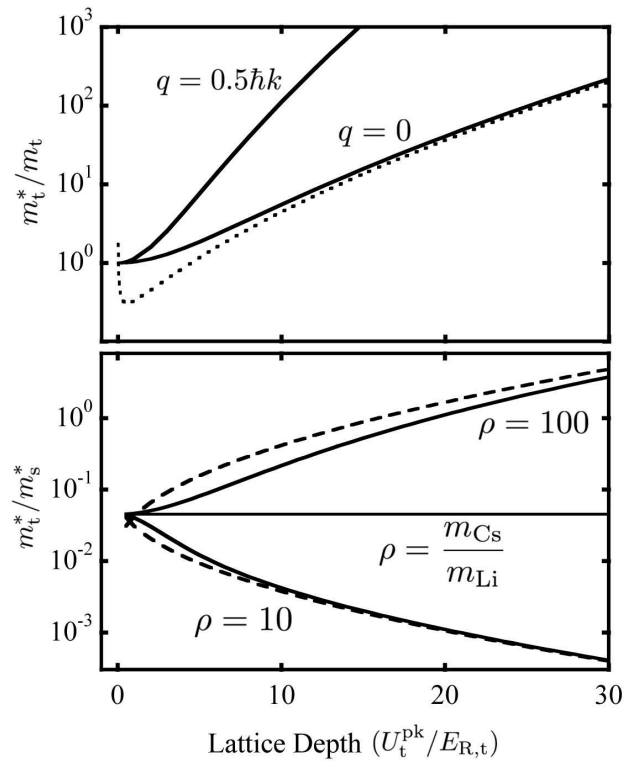}
\caption{{\bf (a)} The ratio of the effective mass to bare mass of the target species is shown as a function of trap depth $\eta_{\rm t}$, for both $q=0$ and $q_\mathrm{c}=0.5 \hbar k$, as labeled. The tight binding approximation (dashed line) approaches the exact calculation for $\eta_{\rm t} \gg 1$. {\bf (b)} The effective mass ratio is shown for the case of a $^{6}$Li target and a $^{133}$Cs spectator. Whether the ratio increase or decreases depends on the selectivity $\rho$ of the lattice. The critical point is when $\rho=m_{\rm s}/m_{\rm t}$ (= 22.2 in this case), as explained in the text. The tight binding approximation of the effective mass ratio is also shown (dashed line).}
\label{fig:em}
\end{figure}
\subsection{Effective mass tuning}

An optical lattice can be used to change the effective mass of the atoms in it, allowing for the tuning of experimental parameters including interaction strength \cite{Bloch2004} and tunneling rate \cite{Meystre2003}, which can be used, for example, to explore different regimes of collective dynamics \cite{Stringari2002}.
The effective mass of a  wave packet centered at quasi-momentum $q_\mathrm{c}$ is 
\begin{equation} \label{eq:em}
m^*(q_\mathrm{c}) = \left[ \frac{d^2 E}{d q^2} \right]_{q_{c}}^{-1},
\end{equation}
where $E(q)$ is the band energy. Figure~\ref{fig:em}(a) shows the effective mass for $q_{{c}}=0$ and $q_{c}=0.5 \hbar k$ in a one-dimensional optical lattice potential $U(x)=U^{pk} \sin^2(k x)$.

For deep lattices, an approximate form for the $q_{c}=0$ case is $m^* = \hbar^2 k^2/2 \pi^2 J$, where $J$ is the tunnelling energy \cite{Zwerger2003}, giving an effective mass enhancement
\begin{equation} \label{eq:mobility}
\frac{m^*}{m} \approx \frac{ e^{2\sqrt{\eta}}}{4 \pi^{3/2} \eta^{3/4}}.
\end{equation}
The ratio of effective masses for the target and spectator can be estimated from Eq.~(\ref{eq:mobility}):
\begin{equation} \label{eq:em_ratio}
\frac{m^*_{t}}{m^*_{s}}  \approx 
\frac{\exp{\left\{ 2\sqrt{\eta_{t}} \left(1 - \sqrt{m_{s}/\rho m_{t}} \right) \right\}  }  } {\rho^{3/4}} \left( \frac{m_{t}}{m_{s}} \right)^{1/4}.
\end{equation}
Figure~\ref{fig:em}(b) shows the effective mass ratio for the case of a $^{6}$Li target and a $^{133}$Cs spectator. In particular, it is striking that for $\rho=10$, the ratio decreases with lattice depth, while for $\rho=100$, the ratio increases with lattice depth. The critical selectivity is well predicted by Eq.~(\ref{eq:em_ratio});  at $\rho=m_{s}/m_{t}$, the effective mass ratio is independent of lattice depth.

Both the tune-out and tune-in schemes provide the means of choosing selectivity. Several tune-in selectivities are shown in Table~\ref{tab:mixtures};  the tune-out selectivity can be chosen simply by choosing a wavelength slightly different from $\lambda_{\rm TO}$. 
In the example used here, the sustainability for $\rho=10$ is better for the tune-in scheme, but tuning to $m^*_{\rm Li} > m^*_{\rm Cs}$ at moderate lattice depths requires a $\rho=100$, for which the tune-out scheme has higher sustainability.


\section {Conclusions}
\label{sec:conclude}

We have discussed how the choice of wavelength used to create an optical lattice can tune its selectivity between elements or isotopes.  This control should increase the range of parameters that can be explored in multispecies ultracold atom experiments.  The tune-out wavelength scheme allows for the complete cancellation of the trapping potential for one species while providing a confining or lattice potential for any other species in the system.  This scheme will work best using the heavier alkali-metal atoms, Rb and Cs, as the spectator elements, and is most successful for the $^{40}$K-$^{133}$Cs fermion-boson mixture.  The alternative tune-in scheme uses a near-detuned optical potential, creating a much stronger potential for one element than the others, without the ability to strictly cancel the potential for one element.  Mixtures involving the Li, Na, and K as spectators are most compatible with this approach, where the $^{40}$K-$^{23}$Na is the most promising fermion-boson mixture.

The power of species selection is in its use to engineer specific lattice environments for the atoms, including adding a bosonic background to mimic the phonons present in solids and tuning the relative effective mass of two species.  
Applications for which experiments are rapid compared to the sustainabilities calculated in Sec.\ \ref{sec:schemes} are especially promising.  Species selection enables  cooling in a two-species mixture, and in the case of fermions trapped in a lattice, reduces the temperature of fermions as they are loaded into a lattice, in contrast to current experimental realizations \cite{Blakie, Kohl2006}. 

\section*{Acknowledgements}
We would like to thank E.\ Arimondo, S.\ Aubin,  M. Extavour, D.\ Jervis, A.\ Mazouchi, J.\ Sipe, and E.\ Taylor for stimulating conversations,  D.\ James for help with atomic calculations, A.\ Paramekanti for help with thermalization calculations, D. Weiss for pointing out the localization advantage of near-detuned blue lattices, and J.\ A.\ Thywissen for suggesting the phrase ``tune-out''.  This work is supported by NSERC and the Canada Research Chairs program.  L.\ J.\ L.\ acknowledges support from NSERC.


\begin{thebibliography}{49}
\bibitem{footnote_2species}{In principle one could use intersection angle to match the periodicity of two-colour standing waves.  In practice, the angle and relative phase would have to exquisitely controlled for cancellation to occur.}
\bibitem{trap_time_footnote}{The sustainability, \textsf{s}, has units of seconds and gives the approximate time it would take to heat the atoms out of the trap.}
\bibitem{poln_footnote} {By ``linear'' polarization, we refer to an equal superposition of
$\sigma^+$ and $\sigma^-$ polarizations.}
\bibitem{footnote_polnaxes}{Unlike for $\sigma^\pm$ or $\pi$, geometric considerations allow that linear polarization can be chosen for all directions of a 3D optical lattice when a weak quantizing magnetic field is present along one of the coordinate axes.}
\bibitem{8ERnote}{In a 3-dimensional optical lattice, the first band is fully confined for lattice depths $\ge 7.4 E_\mathrm{R}$.}
\bibitem{footnote_assume} {We assume that  the species are of equal mass,
that the system is highly degenerate ($T/T_{\rm F} \ll 1$), and that the bosons are either in the condensate or removed immediately from the system when excited.}
\bibitem{footnote1}{1D and 2D lattices may be a way to increase confinement strength
of the first scheme. Assuming high per-site occupancy, the cooling attained
will be governed by single-site physics.}
\end{thebibliography}
\end{document}